\def\plotfiddle#1#2#3#4#5#6#7{\centering \leavevmode
\vbox to#2{\rule{0pt}{#2}}
\includegraphics{#1}}
\def\asr{Adv. Space Res.\rm} 
\def\apj{ApJ\rm} 
 
\def\apjs{ApJS\rm}
\def\mnr{MNRAS\rm}
 
\def\nat{Nature\rm} 
\def\aplc{Astrophys. Lett.+Comm.\rm}  

\def\araa{Ann. Rev. Astron.+Astrophys.}
  
\def\chisq{$\chi^2$} 
\def\chisqr{$\chi^{2}_{\nu}$} 
 
\def\mchisqr{\chi^{2}_{\nu}}

\def\mso{MS1019+5139} 
\def\mst{MS1209+3917}
\def\ros{{\it ROSAT}}
\def\x{\times} 
\def\G{\Gamma} 
\def\a{\alpha} 
 
\def\t{\times} 
\def\funits{\rm ~erg~cm^{-2}s^{-1}} 
\def\lunits{\rm ~erg~s^{-1}} 
\def\nunits{\rm ~cm^{-2}} 
\def\references{\subsection*{REFERENCES} 
\bgroup\parindent=0pt\parskip=\itemsep 
\def\refpar{\par\hangindent=1.2em\hangafter=1}} 
\def\endreferences{\refpar\egroup} 
\def\refpar{\relax} 
\def\reference{\relax\refpar}

\documentstyle{mn}

\begin{document} 

\title{{\bf \ros}~observations of two 'cooling flow' EMSS Galaxies}
 \author[A. J. Blair et al.]
{A.J. Blair, I. Georgantopoulos and G.C. Stewart\\ 
Department of Physics and Astronomy, University of Leicester,   
Leicester, LE1 7RH} 
\maketitle

\begin{abstract} 
We present \ros~ observations of two luminous ($L_x\sim10^{44})
\lunits$ EMSS galaxies, MS1019+5139 and MS1209+3917, previously
classified as 'cooling flow' galaxies.  MS1019+5139 does not appear to
be spatially extended $\rm (<13 $ kpc) while its spectrum is well fit
by a power law with $\rm \G = 1.73 ^{+0.19}_{-0.18}$; X-ray
variability on a timescale of $\sim$ years is also clearly
detected. MS1209+3917 shows no evidence of spatial extension $\rm (<50
$ kpc) but it shows variability, while its spectrum can be fit with
thermal bremsstrahlung emission ($\rm kT=1.8 ^{+0.9}_{-0.4}$ keV) or a
power law model $\rm (\G = 2.50 ^{+0.44}_{-0.42}$, but with excess
photoelectric absorption above the Galactic value).  All the above
argue against thermal emission from a group of galaxies or a galaxy
but in favour of an AGN (possibly BL Lac) interpretation.  We conclude
that no 'normal' galaxies with high X-ray luminosities have yet been
detected in the EMSS survey that could be significant contributors to
the X-ray background.
\end{abstract}


\section{INTRODUCTION}
Although deep \ros~surveys have shown that the majority of soft X-ray
($<2$ keV) sources are type I AGN, i.e. QSOs and Seyfert 1 galaxies
(e.g. Shanks et al. 1991, Georgantopoulos et al. 1996) several
galaxies have also been detected (Boyle et al. 1995, Carballo et
al. 1995, Griffiths et al. 1996).  Their X-ray luminosities ($\rm
L_x\sim 10^{42-42.5} \lunits$) are orders of magnitude higher than
those of nearby galaxies with similar optical magnitudes (Fabbiano
1989). Most of these luminous galaxies have (weak) narrow emission
lines (e.g. Boyle et al. 1995, Griffiths et al. 1996) while few have
the spectra of early-type galaxies, with absorption features only
present (e.g. Griffiths et al. 1995).  It is not clear whether the
X-ray emission is thermal or whether it emanates from an active
nucleus.  Obviously the detection of galaxies with such high X-ray
luminosities has important implications for both our understanding of
the physical processes in galaxies and the origin of the X-ray
background.

Several luminous X-ray galaxies have also been detected in the {\it
EINSTEIN} Extended Medium Sensitivity Survey (EMSS) (Gioia et
al. 1990, Stocke et al. 1991).  New optical spectra from Halpern et
al. (1995) show that most of the objects with $\rm L_x > 10^{42}
\lunits$ are intermediate-type Seyferts (e.g. Sy 1.9).  Another class
of luminous galaxies in the EMSS is that of 'cooling flow' galaxies.
Stocke et al. classify an object as a 'cooling flow' galaxy if the
optical spectrum is similar to the spectra of the dominant galaxies in
cooling flow clusters while there is no associated rich cluster of
galaxies. Their optical spectra contain weak starlight absorption and
strong ($\rm W_{\lambda}\geq 25$ \AA), low ionization emission with
$\rm [O\,\sc{ii}]\gg[O\,\sc{iii}]$.  There are at least four cooling
flow galaxies in the EMSS, all with high X-ray luminosities ($L_x\sim
10^{43-44}$ $\lunits$).  Two of these cooling flow galaxies,
MS1019+5139 and MS1209+3917, appear as point-like objects on Palomar E
POSS plates (Maccacaro et al. 1994). Their V CCD magnitudes are 18.09
and 20.0 respectively (Stocke et al. 1991).  Deep (R$\rm <$23)
observations are also available (Gioia \& Luppino 1994), yet neither
source appears to be associated with a cluster of galaxies.  These
observations show that MS1019+5139 is a point-like source while the
MS1209+3917 appears to be associated with a faint galaxy.  Initial
follow-up observations showed that they both present [O\,{\sc ii}]
lines with equivalent widths $\rm W_{\lambda}> 20\AA$~(Stocke et
al. 1991). $\rm H_{\a}$ was not detected in MS1019+5139 while it was
out of the observed wavelength range in MS1209+3917.  The original
reported redshifts are $\rm z=0.141$ and $\rm z=0.331$ for MS1019+5139
and MS1209+3917 respectively (Stocke et al. 1991).  However, after our
X-ray observations, new spectroscopic observations (Perlman et
al. 1996a and Stocke 1996 priv. comm.) show that the [O\,{\sc ii}]
emission line in MS1019+5139 is quite weak ($\rm W_{\lambda}=14$ \AA)
while $\rm H_{\a}$ and $\rm [N\,{\sc ii}]\lambda\lambda6548,6583$
emission is found with $\rm W_{\lambda}=8\AA$. In the light of the new
optical data, Perlman et al. (1996a) considered a possible BL Lac
interpretation for MS1019+5139, however noting that the presence of
weak emission lines with $\rm W_{\lambda}>5\AA$ would be unique among
X-ray selected BL Lacs.  The new spectroscopic observations (Stocke
1996 priv. comm.) fail to confirm the presence of an oxygen emission
line in MS1209+3917. Instead, absorption lines are detected, which
give a new redshift of $\rm z=0.6$ for this object; we adopt this
value for the redshift of \mst~throughout this paper.  Finally, both
objects were detected at radio wavelengths (5 cm) with the VLA. Their
fluxes are 2.4 and 4.6 mJy respectively (Stocke et al. 1991).  Perlman
et al. (1996a) also present Position Sensitive Proportional Counter
(PSPC) data for \mso~from the \ros~ all-sky survey (Molendi et al. in
preparation). They find a hardness ratio that corresponds to a
power-law energy index of 0.52.  Unfortunately the very low exposure
($\sim 500$ s) does not allow a detailed fit to the spectral data.

Here, we discuss long exposure ($\sim10$ ksec) pointed PSPC and High
Resolution Imager (HRI) observations of these galaxies in an attempt
to constrain the origin of the powerful X-ray emission, and to
investigate possible links between these and the luminous galaxies
detected in deep \ros~surveys. The new PSPC obsevations can provide
useful constraints on the X-ray spectrum of these objects while the
HRI data are important in constraining their size.  Finally, the use
of both the PSPC and the HRI data allows us to investigate the flux
variability over long (year) and short (hour) timescales.

Throughout this work, values of $\rm \rm 50~km~s^{-1}Mpc^{-1}$ and 0.5
are used for $\rm H_0$ and $\rm q_0$ respectively.

\section{The \ros~observations}
MS1019+5139 was observed with the PSPC detector (Pfeffermann et
al. 1986) on board \ros~(Tr\"{u}mper 1983) on 21st October 1993 while
MS1209+3917 was observed on the 3rd June 1994.  After rejecting the
data for times of high particle rates (Master Veto rate $>$ 170,
e.g. Snowden et al. 1992) we end up with exposure times of 7488 and
11857 s for MS1019+5139 and MS1209+3917 respectively.  For the
estimated fluxes and luminosities, see tables~\ref{tab:ms10spec}
and~\ref{tab:ms12spec}. Note that the fluxes and luminosities quoted
throughout the paper are not corrected for absorption.  The source
centroids were estimated using the intensity weighted centroid
algorithm in the \textsc{Starlink Asterix} package. The estimated
celestial coordinates (J2000) are $\rm \alpha=10^h22^m12 \fs 4$,
$\delta=+51^{\circ}24'1 \farcs 0$ for MS1019+5139 and
$\alpha=12^h11^m34 \fs 3$, $\delta=+39^{\circ}00'55 \farcs 0$ for
MS1209+3917, with an error circle radius of $\sim5$ arcsecs. The
proposed optical counterparts are offset from these positions, in the
R.A. and dec. directions, by +2.3, $-0.7$ and $-1.3$, $-0.7$ arcsec
for MS1019+5139 and MS1209+3917 respectively. Positive offsets
indicate optical positions to the North and East of the X-ray
centroids.  A PSPC serendipitous observation of \mst~is also used (ROR
rp700725), obtained on 31st May 1991, having an exposure time of 38125
s, with \mst~at a position $\sim 22$ arcmin off-axis but not obscured
by the PSPC window support structure. These data were retrieved from
the University of Leicester LEDAS \ros~public archive.

\begin{figure}
\plotfiddle{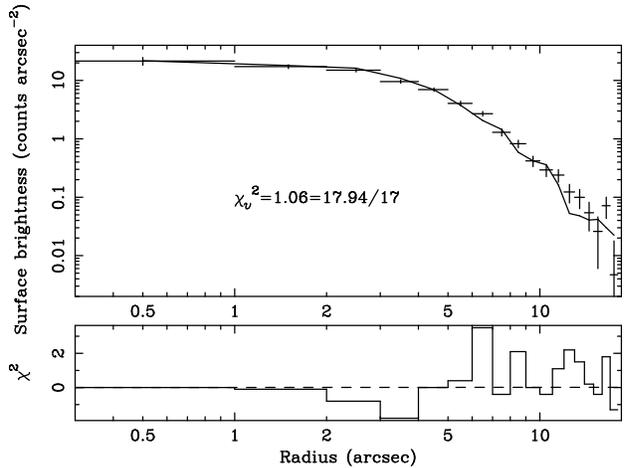}{180pt}{-90}{33}{33}{-130}{200}
\caption{ The HRI Radial profile of MS1019+5139 (with $\rm 1~\sigma$
errors). The profile of AR Lac is shown (solid line).
A 99 per cent upper limit to the linear size of the object was found to be
12.8 kpc (see text for details).  Contributions from each bin to the
\chisq~(from a comparison of the AR Lac data and the \mso~data) are
shown in the lower panel.}
\label{fig:ms10radial}
\end{figure}

Data from two \ros~HRI pointed observations
are used, firstly an on-axis observation of MS1019+5139 (ROR rh600125)
with an exposure of 10924 s, taken between 26th October 1991 and 18th
July 1992 and an observation of \mst~obtained on 18th May 1995 with
an exposure of 11360 s (ROR rh701844).

\section{RESULTS} 

\subsection{MS1019+5139}
In Fig. 1 we present the radial count distribution for MS1019+5139
from the HRI observation.  Data were binned into 1 arcsec bins, out
to, and including, a radius of 18 arcsec, the radius which encloses
$\approx85$ per cent of the source photons.  To determine an upper
limit to the angular size of the object, we modelled the data as a
simple Gaussian distribution of variable sigma.  An upper limit to the
gaussian $\sigma$ is found, at the 99 per cent confidence level, to be
3.45'', corresponding to a linear distance, at the redshift of the
object ($\rm z=0.141$), of 12.8 kpc.  For comparison, the radial count
distribution of an on-axis HRI observation of AR Lac (a point source
used in calibrating the \ros~HRI) is plotted as a solid line.  A
\chisq~value was found from comparison of the \mso~HRI radial profile
and that of AR Lac (normalizing the background subtracted data from AR
Lac to the background subtracted data from \mso), giving
$\mchisqr=1.06$ from 17 degrees of freedom.

We have divided the PSPC and the HRI data into intervals of 1 hour and
400 s respectively, in order to look for short-term variability. The
results are shown in Fig. 2. It can be seen there is no evidence for
short-term variability, however, comparison of the PSPC flux
($2.75\pm0.10 \x 10^{-12}\funits $) with the HRI fluxes ($\rm
2.76\pm0.26\x10^{-12}\funits$ and $\rm 4.88\pm 0.32 \x 10^{-12}
\funits$) and the \ros~all-sky survey flux from Perlman et al. (1996a)
($4.03 \x 10^{-12} \funits$) suggests variability on the long-term
(year) timescale.  The above fluxes refer to the 0.1-2.0 keV band, and
were estimated using the best fit power-law model described below.  A
\chisq~test rejects the null hypothesis of no long-term variability at
the 99.99 per cent confidence level. We note that the observed
variability cannot be due to uncertainties or variations in the
calibration of the PSPC and the HRI.  Hasinger et al. (1993) present
the results of a long term program of calibration observations with
the PSPC, concluding that over a 3 year period, the count rate of the
supernova remnant N132D measured in the PSPC was constant within 2 per
cent.  Similar work using the HRI (David et al. 1996) finds that the
observed count rate of N132D to vary by $\pm5$ per cent over the
period 1990-1995.  We also note that there is an excellent agreement
(better than 1 per cent) between the HRI and PSPC flux calibration.
This is derived by comparing the PSPC and HRI flux of the galaxy
cluster A2256 (David et al. 1996), where the count rates are converted
to flux using the X-ray spectrum from Markevitch \& Vikhlinin (1997).
In contrast, the count rate of MS1019+5139 in the HRI increases by 70
per cent in $\sim6$ months and a comparison of the two PSPC fluxes
shows a 40 per cent decrease between the all-sky survey and the
pointed observation.  We choose not to include in our analysis the
{\it Einstein} IPC measurement obtained about 10 years earlier, due to
differences in the calibration of the responses of the two satellites
(Fiore et al. 1994).

\begin{figure}  
\plotfiddle{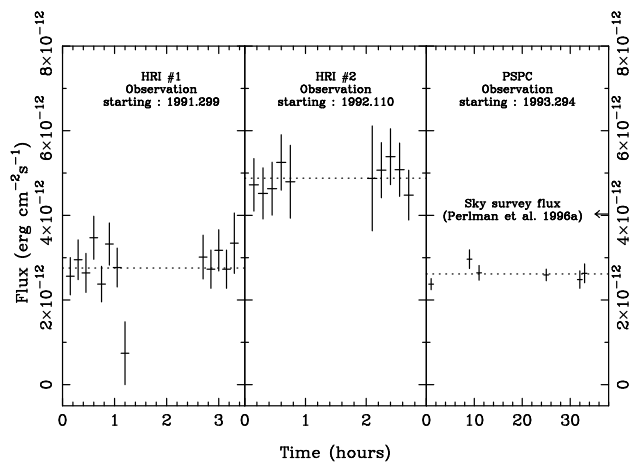}{180pt}{-90}{32}{32}{-130}{180}
\caption{ The variation of source flux with time for MS1019+5139, from
both the PSPC observation (right hand panel) and the two parts of the
HRI observation (left and center panels). The mean fluxes are shown as
dotted lines and the PSPC all-sky survey flux from Perlman et
al. (1996a) is
denoted by the arrow (see text for details).  Fluxes were found in the
energy range 0.1-2.0 keV, and used the best fit power-law model ($\rm
\G=1.73,\, N_H=1.34\x10^{20}\nunits$).  The x-axis shows the time from
the start of each separate observation, in hours. Errors on the data
are 90 per cent errors, from photon statistics and uncertainties in
the model used for converting count rates to fluxes.}
\label{fig:ms10time}
\end{figure}

Spectral fits were made with the simplest possible models: a power law
and a Raymond-Smith hot plasma model (Raymond \& Smith 1977), both
with an absorption component included, having the column density
either fixed to the Galactic value, or as a free parameter.  The
results are presented in table 1. We see that the power law model
provides a marginally better fit compared to the plasma model, albeit
not statistically significant. The value for the F-statistic is 0.818,
implying that a power-law model gives a better fit only at the $\rm
\sim 30$ per cent level.  The derived value for the hydrogen column is
$\rm N_H=1.34 \x 10^{20} \nunits$, with $\rm \G=1.73$, somewhat above,
but within the 90 per cent errors of the Galactic $\rm N_H$ value from
Stark et al. (1992) of $\rm N_H=0.88 \x 10^{20} \nunits$.  The photon
index is consistent, within $\rm \sim2 \sigma$ errors, with the photon
index from Perlman et al. (1996a), of $\rm \G=1.52$.  The folded
spectrum for the best fit power law model, along with the residuals,
is given in Fig. 3.  We find that the best fit temperature of the
Raymond-Smith model is $\rm kT=3.7$ keV using the Galactic absorption
column density and the heavy element abundance fixed at 0.3 solar
units. Allowing the heavy element abundance to be a free parameter, we
find a best-fit value of 0.0 solar units, the 90 per cent upper limit
being 0.2 solar units. The derived temperature is not significantly
changed.

\begin{figure}
\plotfiddle{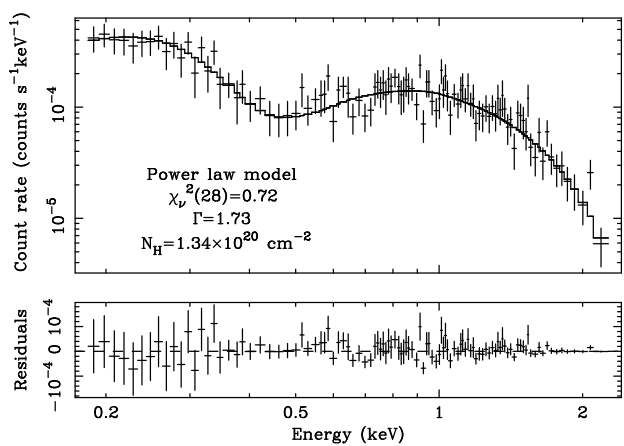}{180pt}{-90}{33}{33}{-130}{180}
\caption{ The spectral data for MS1019+5139, shown with the best
fitting power-law model (solid line) and the residuals from this model
plotted below.  }
\label{fig:ms10spec}
\end{figure}

\subsection{MS1209+3917}
The observed radial count distribution from the HRI data is given in
Fig. 4, together with the radial count distribution found from an
on-axis observation of AR Lac.  Comparison of the \mst~data with the
AR Lac data gives a \chisqr~value of 0.8 for 17 degrees of freedom,
clearly showing that our object is a point source.  Using a similar
method to determine the radial extension as for MS1019+5139, we found
an upper limit of $\sigma=4.06''$, giving a value for the 99 per cent
upper limit to the linear size of the object of 49.5 kpc.


\begin{figure}
\plotfiddle{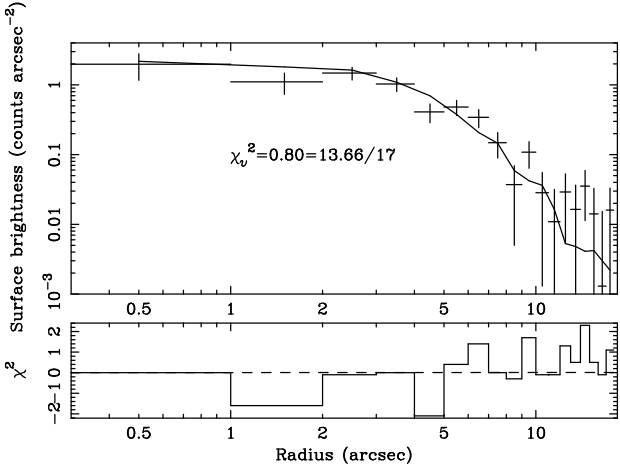}{170pt}{-90}{33}{33}{-130}{180}
\caption{ The HRI radial profile of MS1209+3917. Data are shown with
$1~\sigma$ errors.  The profile of AR Lac is shown as a solid line.
A 99 per cent upper limit to the linear size of the object was found to be
49.5 kpc (see text for details).  In the lower panel are the
contributions to the \chisq~from each bin, caculated from a
comparison of the AR Lac data with the \mst~data.  }
\label{fig:ms12radial}
\end{figure}

Like MS1019+5139, this source is seen to vary on year timescales,
(Fig. 5), evidenced by the fluxes found, in the 0.1-2.0 keV band, from
the pointed PSPC observation, the serendipitous observation and the
HRI data, which are $\rm 6.73 \pm 0.35 \t 10^{-13} \funits, 8.69 \pm
1.61 \t 10^{-13} \funits ~and~ 3.43 \pm 0.31 \t 10^{-13} \funits$
respectively. The conversion from count-rate to flux was performed
using the best-fit thermal model, described below. No significant
variation in flux is seen throughout the duration of each individual
observation.

Spectral fitting was performed using first a power law model and then
a Raymond-Smith plasma model. The results are presented in table 2,
and Fig. 6 shows the folded spectrum with the best fitting plasma
model.

\begin{figure} 
\plotfiddle{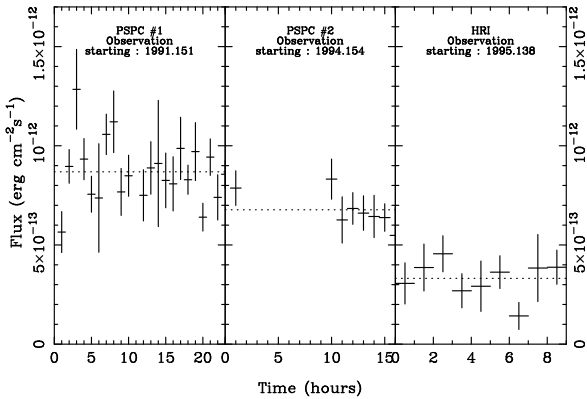}{180pt}{-90}{30}{30}{-130}{180}
\caption{ The variation with time of the flux of MS1209+3917, from the
serendipitous PSPC observation (left hand panel), the pointed PSPC
observation (center panel) and the pointed HRI observation (right hand
panel). The mean fluxes are shown as dotted lines. Count rates were
converted to fluxes in the energy range 0.1-2.0 keV using the best fit
Raymond-Smith model.}
\label{fig:ms12time}
\end{figure}

Both models give acceptable fits when the column density is a free
parameter, although the plasma model does not statistically require an
absorption column greater than the galactic value of $\rm N_H=1.86\x
10^{20} \nunits$, (Stark et al. 1992). However, we find the column
density derived for the power-law model is substantially higher ($\rm
N_H=4.85\x 10^{20} \nunits$) than the observed Galactic value, and
this excess above the Stark et al. value is significant at over the 90
per cent level. The corresponding photon index is $\rm \G=2.50$.  When
we fix the $\rm N_H$ value to the Stark et al. value we obtain a value
of $\rm \G=1.74$ for the power-law index.  The best fit temperature
found with the plasma model with fixed column density and the heavy
element abundance fixed at 0.3 solar units is 1.8 keV. The heavy
element abundance is not well constrained: the best-fit value is 0.56
solar units with an upper and lower limit of 1.8 and 0.2 solar
respectively at the 90 per cent confidence level.

\section{\bf DISCUSSION}

\subsection{MS1019+5139} 
The detection of variability in the \ros~observations
immediately discriminates against a thermal origin for the radiation,
ruling out emission from hot diffuse gas in an early-type galaxy or a
group of galaxies. A hot intergalactic gas origin is further ruled out
by the lack of spatial extension in the HRI observations which
limits the size of the source to $\rm \sim13$ kpc. Hence, the X-ray
data favour an AGN interpretation. 

The absence of strong emission lines in the optical spectrum (weak
[O\,{\sc ii}] and $\rm H_{\alpha}$ + [N\,{\sc ii}] only are present)
argues against a QSO or Seyfert source and could suggest possibly a
galaxy similar to those found in deep \ros~surveys, or alternatively a
BL Lac object. We note however that, the X-ray luminosity of
\mso~($\rm L_x > 10^{44} \lunits$) is around an order of magnitude
brighter than the galaxies found in deep \ros~surveys.  Perlman et
al. (1996a) first suggested a BL Lac identification based mainly on
the optical spectra and the lack of spatial extension in
X-rays. However, they noted that the equivalent widths of the emission
lines are larger than those of X-ray selected BL Lacs. We further note
that the X-ray spectrum derived here is well represented by a
power-law with $\rm \Gamma=1.73^{+0.19}_{-0.18}$, much flatter than
the canonical X-ray selected BL Lac spectrum of $\rm \Gamma\sim2.52$
(Padovani \& Giommi 1996).  The detection of \mso~in radio wavelengths
adds support to a BL Lac scenario, as most X-ray selected BL Lacs show
strong radio emission (Perlman et al. 1996b). Indeed, with values of
1.06 and 0.25 for $\rm \a_{ox} ~and~\a_{ro}$ respectively (Stocke et
al. 1991), the location of this object on the $\rm \a_{ox}-\a_{ro}$
plane (optical to X-ray vs. radio to optical spectral index) is very
close to the X-ray selected BL Lac region, approximately delimited by
the values $\rm 0.6 < \a_{ox} < 1.2 ~and~0.3 < \a_{ro} < 0.5$ (see
Fig. 6 of Stocke et al. 1991). We emphasize however that the X-ray
data alone cannot provide a definitive identification for this object.
Further polarization and radio observations are necessary to prove
that \mso~ is a BL Lac object.

\begin{figure}
\plotfiddle{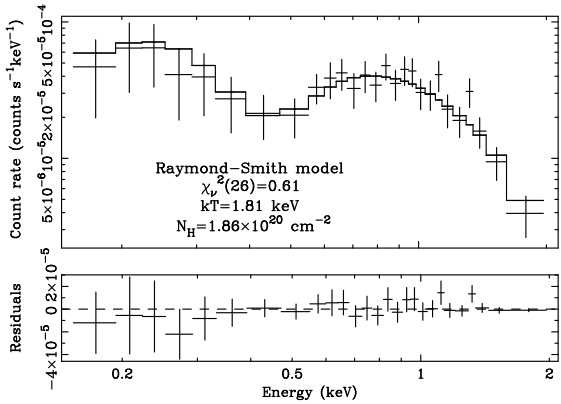}{180pt}{-90}{30}{30}{-130}{180}
\caption{ The observed spectral data for MS1209+3917, with the best
fit Raymond-Smith model and residuals.}
\label{fig:ms12spec}
\end{figure}

\subsection{MS1209+3917}
The lack of X-ray extension strongly argues against thermal emission
from intergalactic gas in a group or cluster of galaxies. The size of
\mst~ is $<$50 kpc at the 99 per cent confidence level, compared to
$\sim$260 kpc for the cluster 0055-279 (Roche et al. 1995), which is
at similar distance (z=0.561) and has a comparable luminosity,
$L_x\approx 10^{44}$ $\lunits$.  Furthermore, MS1209+3917 does not
show any evidence for a cluster on deep CCD images in clear contrast
with the 0055-279 case.  Thermal emission from a compact group of
galaxies is unlikely as the X-ray luminosity of MS1209+3917 is that of
rich clusters of galaxies, while typical group luminosities are
$L_x<10^{43}$ $\lunits$ (e.g. Ebeling, Voges, B\"{o}hringer 1994).
Moreover, the strongest argument against thermal emission comes from
the detection of long-term (year timescale) variability between the
PSPC and the HRI observation.

The X-ray luminosity of \mst~is approximately two orders of magnitude
higher than that of the most luminous nearby, optical galaxies
observed by {\it EINSTEIN} (Fabbiano 1989) and about an order of
magnitude higher than the galaxies detected in deep \ros~ surveys
(Griffiths et al. 1995, Moran et al. 1996). All the above strongly
suggest the presence of an active nucleus. The optical spectrum, which
is unfortunately unpublished yet, holds important clues on the nature
of this object. The absence of emission lines makes a BL Lac
intepretation a strong possibility.  The X-ray spectrum power-law
photon index ($\rm \Gamma=2.50^{+0.44}_{-0.42}$) is similar to the
average 0.1-2.0 keV photon index of X-ray selected BL Lacs
($\Gamma=2.52$) (Padovani \& Giommi 1996). We observe significant
absorption ($\rm N_H= 4.85^{+2.04}_{-1.62} \x 10^{20} \nunits$) above
the Galactic value of $\rm N_H= 1.86 \x 10^{20} \nunits$; such an
excess absorption is not atypical in BL Lacs (Urry et al. 1996). The
BL Lac hypothesis is further supported by the broad band spectral
indices ($\rm \a_{ox}=1.05, \a_{ro}=0.44$), which place \mst~well
within the X-ray selected BL Lac region on the $\rm \a_{ox}-\a_{ro}$
plane.

\section{\bf CONCLUSIONS}
We have presented the results of \ros~observations of two X-ray highly
luminous galaxies (MS1019+5139 and MS1209+3917) from the EMSS,
previously classified as 'cooling flow' galaxies, in an attempt to
constrain the origin of the powerful X-ray emission.  The HRI
observations suggest that neither of the two objects shows evidence of
spatial extension.  The PSPC spectra are consistent with a power-law
model, although a bremsstrahlung model cannot be strongly ruled out.
Finally, long-time variability is clearly detected between the PSPC
and HRI observations in both objects.  The above together with the
high X-ray luminosities of these objects strongly rule out a thermal
emission model from a 'normal' galaxy or a cluster of galaxies and
clearly support an AGN interpretation.  The possibility that there is
some connection to the non-QSO faint sources detected in deep ROSAT
surveys (e.g. Boyle et al. 1995) cannot be definitively ruled out on
the basis of the X-ray data alone.  However, the absence of strong,
broad emission lines in the optical spectra as well as the detection
of radio emission, possibly favours a BL Lac interpretation.  Of
course, further variability and polarization observations in the
optical and radio wavebands are necessary to confirm the BL Lac
hypothesis. We note that if MS1019+5139 is found indeed to be a BL Lac
then it will be an atypical one both because of its flat X-ray
spectrum and its optical emission lines and will be worthy of further
study.

\begin{center}
\begin{table}
\begin{tabular}{|c||c|c|c|c|c|c|} \hline
Spectral& \chisqr($\nu$) &$\rm N_H$&$\rm {\G}_x$\rm &kT&$F_x$\rm &$L_x$ \\
Model & & & & & & \\ \hline
P-L     & 0.72(28) & $1.34^{+0.50}_{-0.45}$ &  $1.73^{+0.19}_{-0.18}$&-& 2.75 & 1.93 \\
P-L     & 0.79(29) & $0.88^{\dagger}$       &  $1.56^{+0.05}_{-0.07}$&-& 2.76 & 1.94 \\
Plasma  & 0.80(28) & $0.69^{+0.21}_{-0.20}$ &-&$4.62^{+2.78}_{-1.32}$  & 2.84 & 1.99 \\
Plasma  & 0.85(29) & $0.88^{\dagger}$       &-&$3.71^{+1.07}_{-0.74}$  & 2.81 & 1.97 \\ \hline
\end{tabular}
\caption{Spectral fitting results for MS1019+5139. The absorption
columns denoted by a dagger are the Galactic values from Stark
et al. (1992) and are in units of $\rm (10^{20} \nunits)$. $\rm F_x$
is  the flux in the energy band 0.1-2.0 keV,
expressed in units of $\rm 10^{-12} \funits$. $\rm L_x$ is the luminosity, in
the same energy band, expressed in units of $\rm 10^{44} \lunits$. The
plasma temperature, kT, is in keV.}
\label{tab:ms10spec}
\end{table}
\end{center}

\begin{center}
\begin{table}
\begin{tabular}{|c||c|c|c|c|c|c|} \hline
Spectral& \chisqr($\rm \nu$) &$\rm N_H$                    &$\rm{\G}_x$&kT&$\rm F_x $&$\rm L_x$ \\
Model& & & & & & \\
\hline
P-L     & 0.68(25) & $4.85^{+2.04}_{-1.62}$ &     $2.50^{+0.44}_{-0.42}$&-&$6.73$&5.08 \\
P-L     & 1.09(26) & $1.86^{\dagger}$       &     $1.74^{+0.14}_{-0.16}$&-&$6.17$&4.66 \\
Plasma  & 0.60(25) & $2.26^{+0.92}_{-0.72}$ & - & $1.71^{+0.78}_{-0.46}$  &$6.28$&4.75 \\ 
Plasma  & 0.61(26) & $1.86^{\dagger}$       & - & $1.81^{+0.90}_{-0.40}$  &$6.37$&4.81 \\ \hline
\end{tabular}
\caption{Spectral fitting results for MS1209+3917. The absorption
columns denoted by a dagger are the Galactic values from Stark
et al. (1992) and are in units of $\rm (10^{20} \nunits)$. $\rm F_x$ is
the flux in the energy band 0.1-2.0 keV,
expressed in units of $\rm 10^{-13} \funits$. $\rm L_x$ is the luminosity, in
the same energy band, expressed in units of $\rm 10^{44} \lunits$. The
plasma temperature, kT, is in keV.}
\label{tab:ms12spec}
\end{table}
\end{center}
 
In summary, the \ros~ observations provide powerful constraints on the 
nature of these objects. Although the exact identification remains 
evasive, we are able to rule out thermal emission models from 
a 'normal' galaxy i.e. a galaxy without an active nucleus. 
We conclude that no 'normal' galaxy  has been detected at high X-ray 
 luminosities, $L_x>10^{43} \lunits$,  in the EMSS or elsewhere, extending 
previous  results by  Halpern et al. (1995).

\section{\bf ACKNOWLEDGEMENTS}
We thank  J. Stocke for communicating details of the optical spectra 
of these objects prior to publication. We  would also like to thank the 
referee for many useful comments which have helped improve this paper. 
 AJB and IG acknowledge University of Leicester and PPARC support
respectively.
This research has  
made use of data obtained through the Leicester Data Archive System 
(LEDAS).

\end{document}